\documentclass[journal=jacsat,manuscript=article]{achemso}

\usepackage[version=3]{mhchem} % Formula subscripts using \ce{}
\usepackage{layouts}
\usepackage{esvect}
\usepackage{siunitx}
\usepackage{amssymb}
\usepackage{booktabs}
\author{Martin Schalk}
\email{martin.schalk@tum.de}
\affiliation{Walter Schottky Institut, Physics Department and MCQST, Technische Universit{\"a}t M{\"u}nchen, 85748 Garching, Germany}
\author{Riccardo Silvioli}
\affiliation{Walter Schottky Institut, Physics Department and MCQST, Technische Universit{\"a}t M{\"u}nchen, 85748 Garching, Germany}
\author{Karina Houska}
\affiliation{Walter Schottky Institut, Physics Department and MCQST, Technische Universit{\"a}t M{\"u}nchen, 85748 Garching, Germany}
\author{Niels van Venrooy}
\affiliation{Walter Schottky Institut, Physics Department and MCQST, Technische Universit{\"a}t M{\"u}nchen, 85748 Garching, Germany}
\author{Katrin Schneider}
\affiliation{Walter Schottky Institut, Physics Department and MCQST, Technische Universit{\"a}t M{\"u}nchen, 85748 Garching, Germany}
\author{Nathan P. Wilson}
\affiliation{Walter Schottky Institut, Physics Department and MCQST, Technische Universit{\"a}t M{\"u}nchen, 85748 Garching, Germany}
\author{Jan Luxa}
\affiliation{Chemistry Department, University of Chemistry and Technology Prague,
16628 Prague, Czech Republic}
\author{Zdeněk Sofer}
\affiliation{Chemistry Department, University of Chemistry and Technology Prague,
16628 Prague, Czech Republic}
\author{Dominik Bucher}
\affiliation{Chemistry Department and MCQST, Technische Universit{\"a}t M{\"u}nchen, 85748 Garching , Germany}
\author{Andreas V. Stier}
\affiliation{Walter Schottky Institut, Physics Department and MCQST, Technische Universit{\"a}t M{\"u}nchen, 85748 Garching, Germany}
\author{Jonathan J. Finley}
\email{finley@wsi.tum.de}
\affiliation{Walter Schottky Institut, Physics Department and MCQST, Technische Universit{\"a}t M{\"u}nchen, 85748 Garching, Germany}

\title[An \textsf{achemso} demo]
{The quantum dynamic range of room temperature spin imaging}
\abbreviations{NV, 2D,}
\begin{document}
\begin{abstract}
\noindent Magnetic resonance imaging of spin systems combines scientific applications in medicine, chemistry and physics. Here, we investigate the pixel-wise coherent quantum dynamics of spins consisting of a 40 by 40 micron sized region of interest implanted with nitrogen vacancy centers (NV) coupled to a nano-magnetic flake of
\ce{CrTe2}. \ce{CrTe2} is an in-plane van der Waals ferromagnet, which we can probe quantitatively by the NV electron's spin signal even at room temperature. First, we combine the nano-scale sample shapes measured by
atomic force microscope with the magnetic resonance imaging data.  We then map out the
coherent dynamics of the colour centers coupled to the van der Waals ferromagnet using
pixel-wise coherent Rabi and Ramsey imaging of the NV sensor layer. 
Next, we fit the pixel-wise solution of the Hamiltonian to the quantum sensor data. Combining data and model, we can explore the detuning range of the spin oscillation with a quantum dynamic range of over $\left|\Delta_{max}\right|=\SI{60}{\mega\hertz}$ in the Ramsey interferometry mode. Finally, we show the effect of the \ce{CrTe2} sample on the coherence of the NV sensor layer and measure a 70 times increase in the maximum frequency of the quantum oscillation going from the Rabi to the Ramsey imaging mode.
\end{abstract}

\section{Introduction}
 Spinful materials, biological reactions and active electronic devices essentially induce magnetic fields that can be imaged on the micron to nano-scale using quantum sensors\cite{schirhaglNitrogenVacancyCentersDiamond2014a,marchioriNanoscaleMagneticField2022, wang2022b}. 
 Image-processed camera read-out and homogeneous quantum control are key challenges in the context of quantum sensing for characterizing and improving smart sensors, coherent memories, bio-molecular and spin devices under ambient conditions.  
Here, we image a layer of nitrogen vacancy centers as quantum sensors coupled to a magnetic van der Waals material to investigate nano-magnetism and the strongly detuned dynamic range of coherent spin oscillations with a wide field of view.
We define the quantum dynamic range of the quantum sensor as the maximum detuning range we can resolve by means of a pixel-wise coherent spin oscillation driven by single frequency pulses. Complementary to the common low quantum dynamic range approaches of optimizing the sensitivity at a single frequency\cite{wolf2015, schmitt2017} and three level atom sensing (3LS) schemes such as spins driven by two frequency tones\cite{bauch2018a, hart2021a, kazi2021, wang2022}, we here explore the full quantum dynamic range of room temperature spin imaging in a strongly detuned magnetic environment.
Magnetic van der Waals materials\cite{yangVanWaalsMagnets2021} like \ce{CrTe2} offer opportunities to investigate magnetic phase transitions, fingerprints of quantum spin phases and are the material of choice to explore the atomic limit of scaling spintronic memories\cite{Yang2022}.
While single spin scanning techniques have been used to image the static magnetization\cite{song2021, thiel2019, fabre2021} and spatially resolved domain walls\cite{tetienne2015, sun2021} of 2D magnets, the wide-field imaging mode\cite{scholten2021} has several benefits: First, the NV ensemble signal is bright and homogeneous and consequently offers higher combined spin sensitivity than using single spins\cite{clevenson2015}.
\begin{figure}[t]
	\centering
	\includegraphics[width=1.0\textwidth, height=0.618\textwidth]{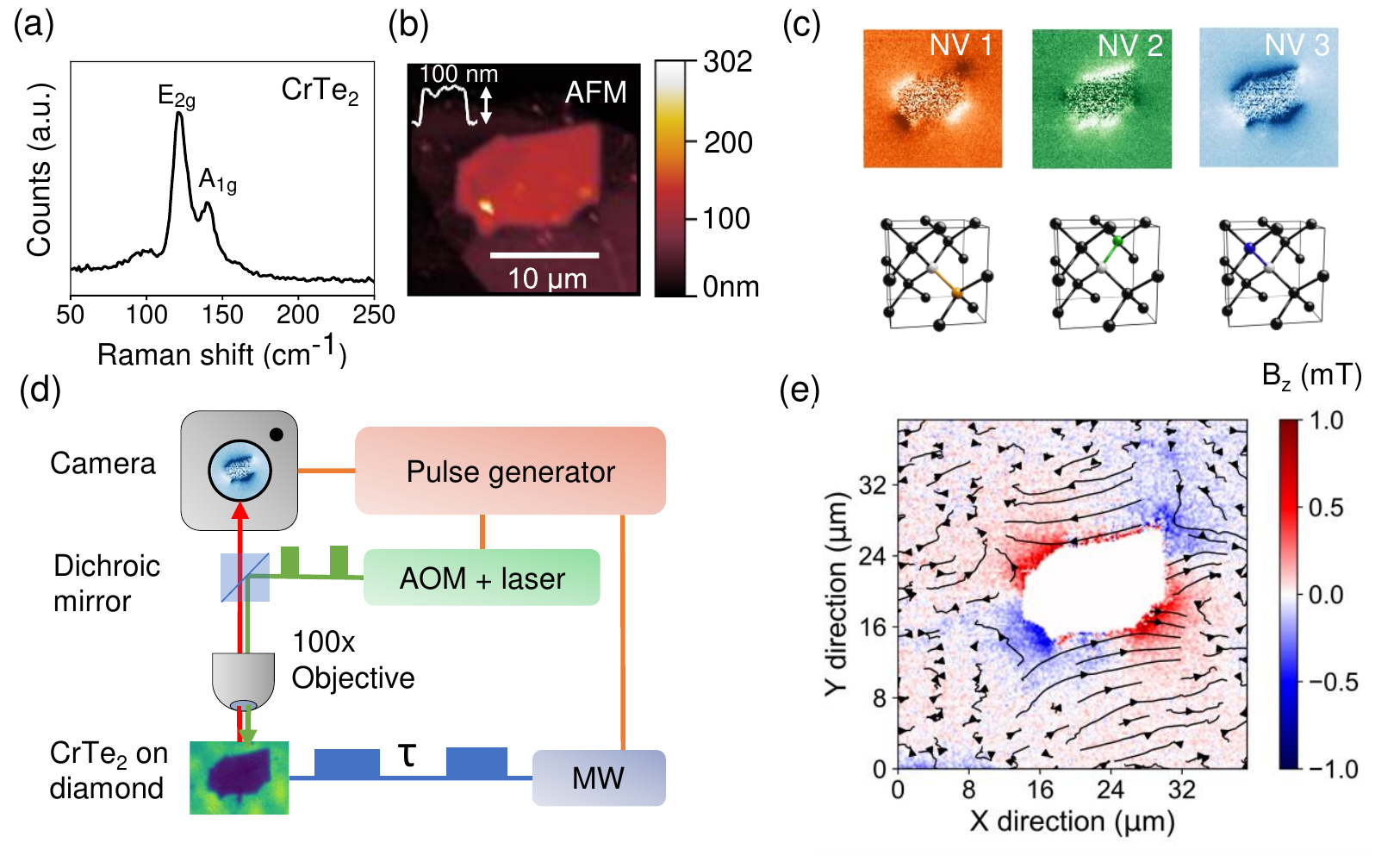}
	\caption{ Quantitative nano-magnetic imaging with a wide-field of view  a) Van der Waals magnet \ce{CrTe} Raman modes (b) Atomic force microscope height profile, inset  horizontal height cut of the \SI{100}{\nano\meter} thick \ce{CrTe2} magnetic flake c) NV spin directions in the diamond lattice used for 3D vector magnetometry d) Coherent wide-field camera setup for direct spin imaging e) Magnetic flux density of \ce{CrTe2} imaged as a quadrupol by the NV electron spin resonance detuning, the colormap depicts the $B_z$ component; $B_x$ and $B_y$ are shown as field lines in black.}
	\label{fig:setup}
\end{figure}
Second, the camera acquires pixel-wise spin information in parallel and thus opens up easy access to imaging science and even real-time image processing on micro-second (camera limited) time-scales\cite{leoni2021}.
Third, the sensor signal is robust and does not bleach or change during subsequent measurements due to the solid-state integration of the quantum sensor layer into the protecting diamond lattice. The point-wise coupling approach in scanning mode measurements\cite{thiel2019, song2021} induces oscillations of the the fork with respect to the sample and microwave field leading to more involved stability and fork oscillation calibration requirements\cite{huxter2022}.
Most importantly, the fixed coupling constants and distances make homogeneous coherent control of the whole NV layer as a fully quantum-enabled sensor with high dynamic range feasible as we outline in this letter.  
\section{Results and discussion}
Figure \ref{fig:setup} schematically portrays the coherent wide-field imaging experiment. 
The synthesized bulk crystals show the Raman $E_{2g}$ and $A_{1g}$ modes of \ce{CrTe2}, verifying the crystallography of the in-plane room temperature ferromagnet\cite{fabre2021}, as measured by Raman spectroscopy in figure \ref{fig:setup} (a). The shape and height of the exfoliated van der Waals flake is mapped out by an atomic force microscope height profile in figure \ref{fig:setup} (b). 
\ce{CrTe2} transferred on top of the diamond sensor induces a magnetic field, which is commonly probed by electron spin resonance (ESR) spectroscopy\cite{balasubramanian2008, maze2008}. The NV ensemble's electron spin resonances are linearly magnetic bias field dependent and therefore naturally calibrated via\cite{steinert2010, maertzVectorMagneticField2010}:
\begin{equation}
\label{eq1}
\begin{split}
\mathcal{H}_{NV,i} &= \gamma_{el} \mathbf{B} \cdot \mathbf{S} + \mathbf{S} \cdot \mathbf{D} \cdot \mathbf{S} \\ & \approx \gamma_{el} B_i\left(\text{sin}\text{\hspace{0.17em}}{\theta }_{B}\cdot \text{cos}\text{\hspace{0.17em}}{\phi }_{B}\cdot {\hat{S}}_{x}+\text{sin}\text{\hspace{0.17em}}{\theta }_{B}\cdot \text{sin}\text{\hspace{0.17em}}{\phi }_{B}\cdot {\hat{S}}_{y}+\text{cos}\text{\hspace{0.17em}}{\phi }_{B}\cdot {\hat{S}}_{z}\right)+ \frac{D}{h}\hat{S}_z^{2} ,\\
    \end{split}
\end{equation}
where $\mathbf{S}$ (S = 1) is the spin operator of the NV electron spin, the index $i$ stands for the eight possible spin orientations (four NV directions in diamond times two spin orientations), $\gamma_{el} = \SI{28}{\giga\hertz\per\tesla}$ is the NV electron's gyromagnetic ratio and the offset $D/h\approx\SI{2.87}{GHz}$ is the zero field splitting. 
%The Eigenenergy relations also define the sensor's natural calibration since the spin resonance energies are to first order tuned by only the local magnetic Zeeman field $B_i$ pointing along the spin directions and natural constants.
%An optically detected magnetic resonance spectrum can be measured at every pixel with resonance energy $E_{NV,i} = hf_{NV,i}$. 
We use three $i=(1,2,3)$ NV spin orientations depicted in orange, green, and blue in figure \ref{fig:setup} (c) to calculate a three-dimensional magnetic field map. Here, each color map represents the measured electron spin resonance frequency shift $\Delta f_{NV,i} \propto \gamma_{el} {\Delta B_i}$. The spin resonance shifts of the three different NV directions directly relate to a 3D cartesian lab coordinate system ($B_x, B_y, B_z$), where the directional expansion in equation \ref{eq1} is determined by the diamond crystal orientation (100) and vector projections.\cite{steinert2010, maertzVectorMagneticField2010, munzhuber2020, broadway2020a}
\begin{figure}[t]
	\centering
	\includegraphics[width=1.0\textwidth, height=0.618\textwidth]{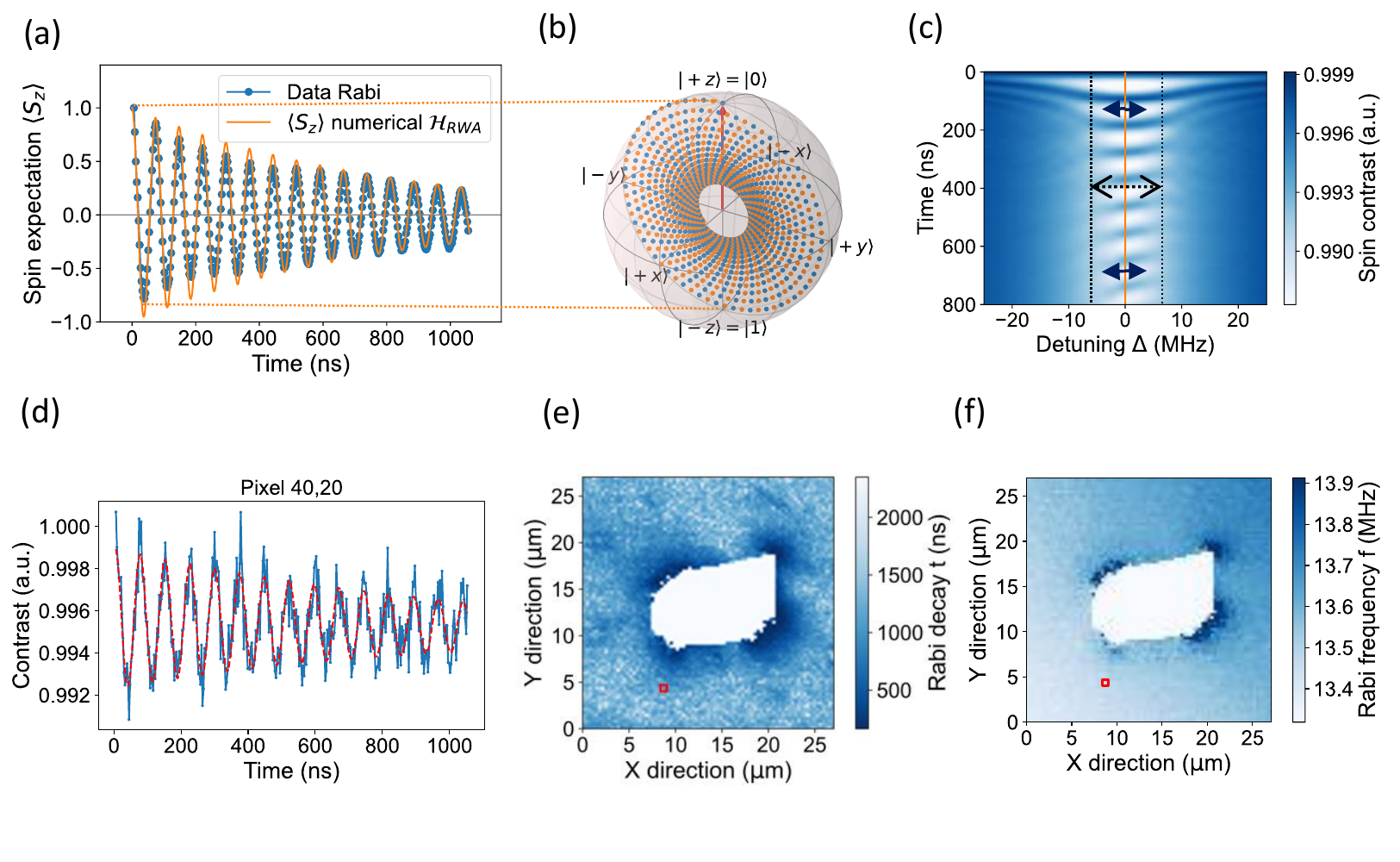}
	\caption{Dynamic range of a pixel-wise detuned Rabi oscillation a) Homogeneous wide-field Rabi spin driving on the bare sensor with Rabi data in direct comparison to the numerical rotating wave approximation model b) Bloch sphere representation of the measured Rabi oscillation c) Measurement of the detuned Rabi oscillation showing a symmetric increase of the Rabi frequency around zero detuning, quantum dynamic range indicated by dashed lines and arrows, sample effect by dark blue arrows
	d) Single camera pixel Rabi oscillation measurement and fit (red dashed line)
	e) Full Rabi decay map, red square indicates pixel (40,20) f) Rabi frequency image with dynamical spin speed-up induced by the \ce{CrTe2} magnet.}
	\label{fig:pixcoherence}
\end{figure}
The 3D magnetic field plot - as color plot for $B_z$ and black field lines for the $B_x$ and $B_y$ components -  in figure \ref{fig:setup} (e)  shows a quadrupolar magnetic field distribution $\mathbf{B}$. Using the NV ensemble spin resonance, we can thus directly image the magnetic flux distribution by the in-plane magnetization of the nanoscopic \ce{CrTe2} sample\cite{fabre2021, kumar2022} at a distance of $z = \SI{100}{nm}$. We combine the shape of the flake as a mask for clarity in the upcoming measurements, since the \ce{CrTe2} sample darkens the NV center photoluminescence signal out in the imaging path.
%Here, \ce{CrTe2} shows a quadrupolar magnetic flux distribution $\mathbf{B}$ induced at the NV sensor plane, going beyond single domain or dipolar magnetic flakes probed earlier\cite{fabre2021a}. 
Additional to spin-mixture induced frequency shifts in continuous wave spin resonance spectroscopy\cite{balasubramanian2019}, the magnetic field distribution of our sample provides a bias field to the NV spin dynamics. 
A triggered camera read-out combined with drive pulses observe spin dynamics of the sensor NV layer coupled to the \ce{CrTe2}. Thus, we can in particular use the coherent dynamics of the NV layer as a tool to probe nano-magnetic samples like \ce{CrTe2} as follows.
The wide-field quantum sensor dynamics $S_{NV,\square}(t)$ is influenced by the three main components drive $\omega_d$, detuning $\Delta$ and decay $\xi_{env}$.
Briefly, the driven NV ensemble resonance can be described as an ensemble of atoms, a quantum two-level system (2LS) with spin-1, \cite{cohen-tannoudji.2010} and spin transition frequency $\omega_{s}$ using the rotating wave approximation:
\begin{equation}
    \mathcal{H}_{RWA}=\frac{\Delta}{2} S_z + 2\eta S_x,
\end{equation}
where $\Delta = \omega_{s,\square } - \omega_{d}$ is the detuning of the spin resonance with respect to  the drive frequency $\omega_{d}$.
As shown in equation \ref{eq1} and figure \ref{fig:setup} (c), the spin transition frequency $\omega_{s,\square } = \frac{f_{NV}}{2\pi} \approx D/h+\gamma_{el} \cdot B_z$ depends for every pixel on the zero field splitting and the local magnetic bias field. 
%The sensor part of a single pixel $\square$ with $n$ NVs is then $H_{\square NV} = \frac{\Delta}{2}\sum_n\sigma_n^z$ with detuning $\Delta = \omega_{s,\square } - \omega_{d}$ where $\omega_{d}$ is the drive frequency.
We initialize the complete sensor plane with a green laser pulse in the ground state and apply a resonant microwave drive $\mathcal{H}_d =  \gamma_{el} b_1 \cos{\left(\omega_s t\right)} S_x = 2\eta S_x  $ with strength $\eta$ to the NV ensemble. With that, we can perform pixel-wise Rabi oscillations with the effective Rabi frequency $\Omega_{\square} = \sqrt{\Delta_{}^{2} + 4\eta^{2}}$.
We use a microwave split-ring resonator antenna\cite{sasaki2016} to apply a homogeneous drive field so that the drive strength $\eta \approx \mathsf{const.}$  $\forall \square$ is constant or continuously defined for all pixels.
Figure \ref{fig:pixcoherence} (a) shows the bare sensor Rabi oscillation and its coherence, where all pixels are averaged into one curve proving the wide-field homogeneity and displaying the good control parameter range for the rotating wave approximation shown on the Bloch sphere \ref{fig:pixcoherence} (b). The orange line shows the numerical solution to the rotating wave approximation model $\mathcal{H}_{RWA}$ fitting well to the wide-field data.
Figure 2c then shows the measured detuned Rabi oscillation following closely the solution $S_{NV,\square}(t) = \sqrt{\frac{\Delta^{2} \sin^{2}{\left(\frac{\Omega t}{2} \right)}}{\Omega^{2}} + \cos^{2}{\left(\frac{\Omega t}{2} \right)}}$ with a dynamic range of $\left|\Delta_{Rabi}\right| = \SI{15}{\mega\hertz}$.
We emphasize, that if both the strength $\eta$ of the microwave field drive and the detuned spin frequency $\omega_{s,\square}$ influence the pixel's Rabi frequency\cite{balasubramanian2019, zhang2021a}, it is difficult to distinguish between sensor or sample-related frequency shifts induced by either inhomogeneous control or magnetic bias.
The Rabi imaging in figure \ref{fig:pixcoherence} shows that the Rabi oscillations (f) get faster and also decay faster (e) for higher magnetic field detuning induced by the \ce{CrTe2} $\Delta_{\ce{CrTe2}}> 0$ as indicated by the blue arrows in (c). The sensor has a slight radial background stemming from the ring resonator's microwave field profile which can be fitted by the plane $\SI{7.94}{\kilo\hertz\per\micro\meter}\cdot X + \SI{6.50}{\kilo\hertz\per\micro\meter}\cdot Y +  \SI{13.38250}{\mega\hertz} = Z$ around the mean Rabi frequency $\Omega=\SI{13.3825}{\mega\hertz}$. In the decay map in figure \ref{fig:pixcoherence} (e) we can clearly identify the magnetic \ce{CrTe2} van der Waals flake by faster relaxation and smaller additional surface background effects which can originate from charged surface contamination or strain, which change the T1 spin-lattice relaxation time\cite{bluvstein2019}.
The pixel-wise Rabi imaging (see for instance pixel 40,20 in (a)) with homogenous wide-field control can to this end be used as a naturally calibrated sensor for temperature, strain, microwave and magnetic gradient sensing.
Subsequently, we investigate Ramsey interferometry and the free induction decay influenced by \ce{CrTe2}. 
\begin{figure}[t]
	\centering
	\includegraphics[width=1.0\textwidth, height=0.618\textwidth]{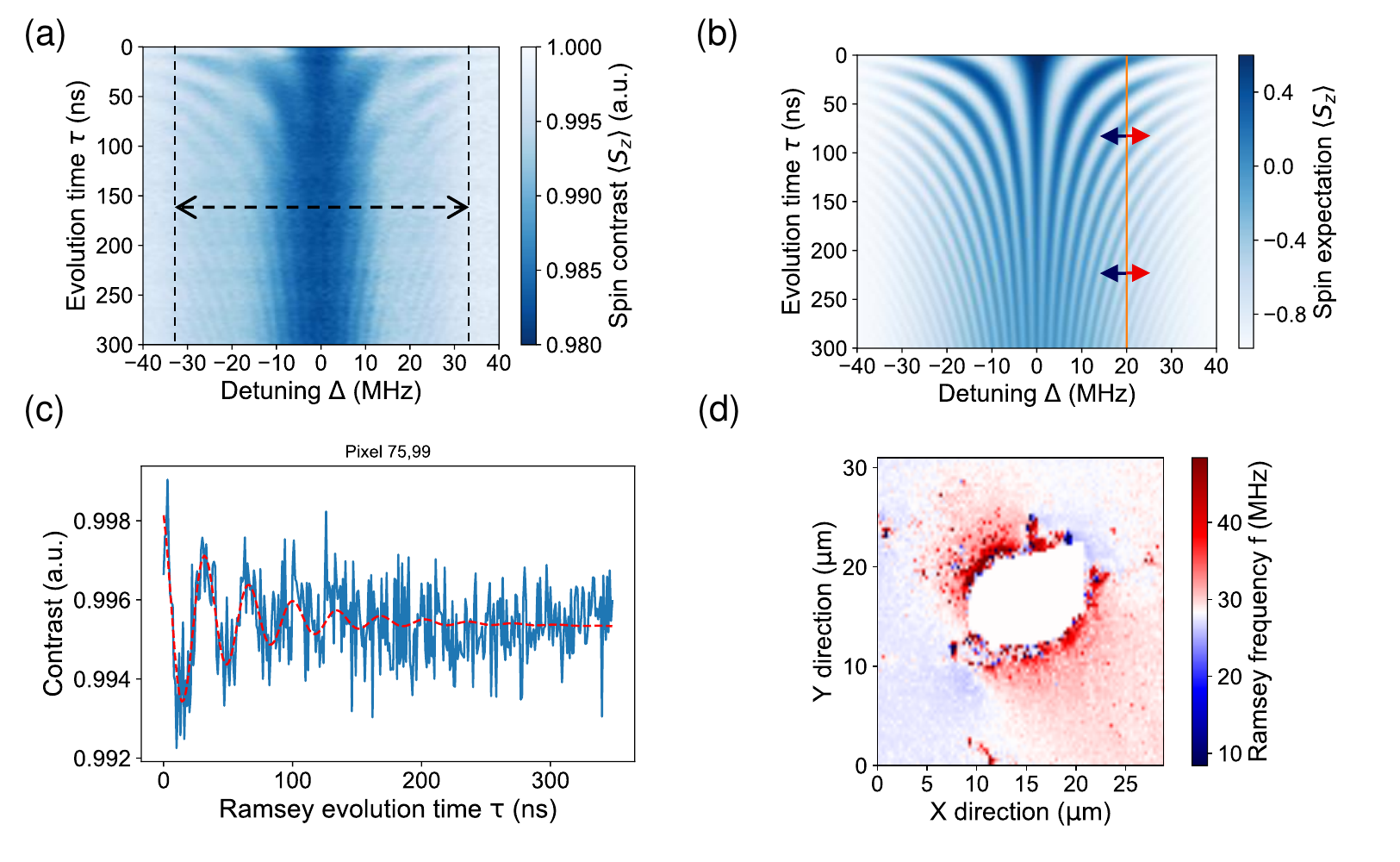}
	\caption{Pixel-wise coherent Ramsey interferometry and detuning a) Ramsey fringes of the whole sensor with detuned microwave drive exploring the full quantum dynamic range marked by dashed lines b) Simulation of the Ramsey fringes using a two-level quantum interference model, the orange line shows the experimental MW drive set-point and the red (blue) arrows indicate the magnetic detuning induced by the sample in positive (negative) direction c) Single pixel Ramsey oscillation data blue and red-dashed trend line fit d) Map of the pixel-wise Ramsey frequency detuned by the \ce{CrTe2} van der Waals magnet.}
	\label{fig:ramsey}
\end{figure}
 The free induction decay of the ensemble is then detected as Ramsey fringes approaching the ideal form (at strong drive $\eta \gg \Delta$):
\begin{equation}
S_{NV,\square}(\tau) =  \sum_n\langle\sigma_{z,n}\rangle\propto \left( \frac{\cos{\left(\Delta \tau \right)}}{2} + \frac{1}{2} \right) \cdot e^{-\tau/T_2^*}  = \cos^2(\Delta\cdot\tau/2) \cdot e^{-\tau/T_2^*}.
\end{equation}
Figure \ref{fig:ramsey} (a) shows the measured quantum interference of the whole sensor chip under detuning. The model in \ref{fig:ramsey} (b) of the Ramsey interferometry assuming a two level quantum interference corresponds  remarkably well with no free parameters to the measurement even in the case that all spins in all pixels evolve and are measured in parallel.
In the following, we drive the spin resonance $f_{res}=\SI{2.3636}{\giga\hertz}$ with drive frequency $f_{drive}=\SI{2.3836}{\giga\hertz}$ at detuning $\Delta=\SI{20}{\mega\hertz}$ as marked in figure \ref{fig:ramsey} (b).
The single pixel (40,20) data in figure \ref{fig:ramsey} (c) shows that a Ramsey oscillation can even be resolved on a single camera pixel without binning.
The Ramsey frequency is then plotted in figure \ref{fig:ramsey} (d) showing a strong quadrupolar detuning of the Ramsey frequency influence by the \ce{CrTe2} sample. The sample speeds up the Ramsey oscillation in the positive magnetically detuned part and decreases Ramsey frequency at the sample edges of negative magnetic field. 
Here, the sample shifts in positive and negative directions and the sign is given by the orientation of the rotating frame defined by the Z spin ensemble direction and the X drive axis.
Ramsey interferometry is sensitive to smaller magnetic field changes and detunings since the change in frequency for the Ramsey measurement $\Delta f_{Ramsey, max} \approx \SI{35}{\mega\hertz}$  is nearly two orders of magnitude higher than the Rabi frequency change $\Delta f_{Rabi,max} \approx \SI{0.5}{\mega\hertz}$ by directly quantitatively comparing the colorbar of the detuned Rabi and Ramsey interferometry measurements on the same sample with  fixed maximum drive power $\eta$. The optimal sensor set-point is centered around the maximum rate of change under detuning $\max \left(\frac{\partial S_{NV, \square} (t)}{\partial \Delta} \right)$, which then equivalently corresponds to the most sensitive quantum dynamic range. The quantum dynamics trend of the increased Ramsey frequency range is also in accordance with Ramsey limited sensing outperforming continuous wave or Rabi protocols\cite{Barry2020}. The quantum dynamic range is over $\left|\Delta_{max}\right|= \SI{60}{\mega\hertz}$  for Ramsey and around $\left|\Delta_{Rabi}\right| = \SI{15}{\mega\hertz}$ for the Rabi imaging mode as marked in figure \ref{fig:ramsey} (a) respectively \ref{fig:pixcoherence} (c).
\begin{table}[t]
\centering
\begin{minipage}{\linewidth}
\begin{tabular}{lcccccl}
\toprule
Technique   & Dynamics & Drive  & Sensitivity & Dynamic Range   & Ref. \\\midrule
Spin Resonance    & spin mixture\footnote{no full coherent oscillation but statistical spin phase mixture} & broad-band  & $<\si{kHz}$ & $>\SI{10}{\giga\hertz}$ & \cite{balasubramanian2008, maze2008}\\
Relaxometry & spin decay  & single f & $<\si{kHz}$ & $\mathcal{O}(\SI{1}{\giga\hertz})$&  \cite{huang2021}\\
Rabi & 2LS, driven & single f  & $<\si{kHz}$ & $\mathcal{O}(\SI{10}{\mega\hertz})$& \cite{mariani2020} \\
Ramsey & 2LS, free & single f  & $<\si{Hz}$ & $\mathcal{O}(\SI{100}{\mega\hertz})$ & \cite{hart2021a, kazi2021}\\
Homodyne & 2LS, synced  & single f  & $<\SI{1}{\milli\hertz}$\footnote{determined spot-wise, not by wide-field imaging, divide by $\gamma$ for magnetic units in $\si{T}$}& $<\SI{100}{\kilo\hertz}$&\cite{schmitt2017, lang2019, wang2022a}\\
Heterodyne  & 3LS & two f/chirp\footnote{smooth transition between multilevel chirp and broad-band spin resonance spectroscopy\cite{niemeyer2013}}  & $<\SI{100}{\milli\hertz}$ & $>\SI{100}{\mega\hertz}$ & \cite{wang2022}\\
\\\bottomrule
\caption{Design aspects and quantum dynamics of spin imaging techniques.}
\label{tab:quantum_dyn}
\end{tabular}
\end{minipage}
\end{table}
Table 1 summarizes the design aspects of quantum dynamic imaging showing that the detection of fast quantum oscillations in a strongly detuned magnetic environment requires a high dynamic sensing range. Therefore, \ce{CrTe2} can be imaged by spin resonance, Rabi and Ramsey characterized in this letter, where Ramsey provides the best compromise in terms of drive-free, sample induced quantum oscillations, a high dynamic range and sufficient sensitivity.
%Approaching the ultimate sensing limit of the setup is requiring laser and microwave excitation powers which can deteriorate the magnetic van der Waals samples. 
Future  homodyne\cite{zhang2021a, wang2022a} or heterodyne\cite{wang2022} sensing protocols together with optimized light collection\cite{allert2022} are clear ways to approach the wide-field sensitivity limit for fast cameras\cite{wojciechowski2018}, but will also either reduce the accessible dynamic range or conceal quantum dynamics in multi-level, incoherent excitations\cite{bauch2018a, hart2021a, kazi2021}.
Here, we emphasize the importance of homogenous wide-field quantum control and coherent, collective spin dynamics investigating a strongly detuned spin environment induced by the \ce{CrTe2} van der Waals magnet. 
Most quantum phase or entanglement-based  sensing schemes then equally require homogeneity and quantum-transduced frequency stability\cite{zhang2022,ebel2021} over the whole sensor plane as an essential prerequisite. 
We conclude that the stable control of the pixel-wise quantum dynamics and a high dynamic detuning range are important design and characterization aspects for practical quantum imaging which must not be omitted next to the optimization of the single spot resolution or the highest sensitivity at low quantum dynamic range.
\section{Experimental}
The spin imaging setup consists of: 
Green laser: DPSS Spectra Physics 532nm at P=0.2W, 
Acousto-optic modulator: G\&H 3350-199, 
Dichroic mirror: Thorlabs DMLP550, 
Pulse generator: Swabian instrument pulsestreamer, 
Time of flight camera: Basler acA 1920-155um, 
Objective: Leica 100x NA 0.75 HC PL FLUOTAR,
Microwave generator: Windfreak SynthHD PRO,
Microwave antenna: Split ring resonator antenna with a homogenous Bz field of \SI{1}{mm} radius centered around \SI{2.8}{\giga\hertz}, (design files are available in the supplementary information).
Sample preparation: element6 electronic grade diamond, implantation Innovion with \ce{^{15}N} fluence \SI{2e12}{\per\centi\meter\squared}
at energy \SI{20}{keV}, diamond is excited from the side in a total internal reflection geometry (cf. supplementary).

\begin{acknowledgement}
All authors gratefully acknowledge the German Science Foundation (DFG) for financial support, as well as the clusters of excellence MCQST (EXS-2111) and e-conversion (EXS-2089).
J. J. Finley acknowledge the state of Bavaria via the One Munich Strategy and Munich Quantum Valley.
We greatfully acknowledge support and discussions to the sample preparation with Bucher lab at TUM chemistry department and Sofer lab at UCT Prague.
\end{acknowledgement}

\begin{suppinfo}
The theoretical models are based on the two-level systems in rotating wave approximation and Ramsey spectroscopy using the quantum optics package\cite{kramer2018}.
All experimental procedures, the RF design files, theoretical models, \ce{CrTe2} synthesis and characterization data for the samples and measurements are available online or upon request.
\end{suppinfo}


%\bibliography{NVmagnet}

\providecommand{\latin}[1]{#1}
\makeatletter
\providecommand{\doi}
  {\begingroup\let\do\@makeother\dospecials
  \catcode`\{=1 \catcode`\}=2 \doi@aux}
\providecommand{\doi@aux}[1]{\endgroup\texttt{#1}}
\makeatother
\providecommand*\mcitethebibliography{\thebibliography}
\csname @ifundefined\endcsname{endmcitethebibliography}
  {\let\endmcitethebibliography\endthebibliography}{}

\end{document}